\documentclass[%
 reprint,preprintnumbers,
nofootinbib,
 amsmath,amssymb,
 aps
]{revtex4-1}
\pdfoutput=1
\usepackage{graphicx}
\usepackage[utf8]{inputenc}
\usepackage{flushend}
\usepackage{dcolumn}
\usepackage{bm}
\usepackage{balance}

\usepackage[normalem]{ulem}

\usepackage[colorlinks = true,
            linkcolor = blue,
            urlcolor  = blue,
            citecolor =green,
            anchorcolor = blue]{hyperref}
\usepackage{verbatim}
\usepackage{color,ulem}
\usepackage[english]{babel}

\newcounter{defcounter}
\newenvironment{myequation}{%
\addtocounter{equation}{-1}
\refstepcounter{defcounter}

\begin{equation}}
{\end{equation}}

\usepackage[utf8]{inputenc}
\input Starburst.fd
\newcommand*\initfamily{\usefont{U}{Starburst}{xl}{n}}\initfamily

\newcommand{\beq}{\begin{eqnarray}}
\newcommand{\eeq}{\end{eqnarray}}
\usepackage{amsmath}
\usepackage{tikz}
\usetikzlibrary{decorations.pathmorphing}
\usetikzlibrary{shapes.misc}
\tikzset{cross/.style={cross out, draw=black, minimum size=8*(#1-\pgflinewidth), inner sep=0pt, outer sep=0pt},
cross/.default={1pt}}
\usetikzlibrary{patterns,math}
\begin{document}

\preprint{IFT-UAM/CSIC-20-139}

\title{\Large  How small  hydrodynamics can go}

\author{\textbf{Matteo Baggioli} \vspace{0.2cm}}%
 \email{matteo.baggioli@uam.es}
 
 \vspace{1cm}
 
 \affiliation{Instituto de Fisica Teorica UAM/CSIC, c/Nicolas Cabrera 13-15,
Universidad Autonoma de Madrid, Cantoblanco, 28049 Madrid, Spain.\vspace{0.2cm}}
\affiliation{Wilczek Quantum Center, School of Physics and Astronomy, Shanghai Jiao Tong University, Shanghai 200240, China}
\affiliation{Shanghai Research Center for Quantum Sciences, Shanghai 201315.}

\begin{abstract}
Numerous experimental and theoretical results in liquids and plasmas suggest the presence of a critical momentum at which the shear diffusion mode collides with a non-hydrodynamic relaxation mode, giving rise to propagating shear waves. This phenomenon, labelled as ''k-gap'', could explain the surprising identification of a low-frequency elastic behaviour in confined liquids. More recently, a formal study of the perturbative hydrodynamic expansion showed that critical points in complex space, such as the aforementioned k-gap, determine the radius of convergence of linear hydrodynamics -- its regime of applicability. In this work, we combine the two new concepts and we study the radius of convergence of linear hydrodynamics in ''real liquids'' by using several data from simulations and experiments. We generically show that the radius of convergence increases with temperature and it surprisingly decreases with the \color{black} electromagnetic \color{black} interactions coupling. More importantly, \color{black}for all the systems considered\color{black}, we find that such radius is set by the Wigner–Seitz radius -- the characteristic inter-atomic distance of the liquid, which provides a natural microscopic bound.

\end{abstract}

\maketitle
\subsection*{Introduction}
''$\Pi \acute{\alpha} \nu \tau \alpha\,\,\rho \epsilon\iota$'' -- everything flows. Hydrodynamics is an effective field theory (EFT) formulated as a perturbative expansion in spatial and time gradients. It governs the dynamics of conserved quantities which in Fourier space can be constructed as an infinite expansion in frequency $\omega$ and momentum $k$ -- from slow processes and large scales, to fast dynamics and short lengths. It applies to the most disparate systems, from liquids \cite{landau2013fluid} and solids \cite{PhysRevA.6.2401} to flocks \cite{TONER2005170}, crowds \cite{Bain46} and even to financial markets \cite{PhysRevLett.112.098703}. As every EFT, the microscopic physics is hidden in an infinite set of unknown coefficients since it is ''irrelevant'' (in the Renormalization Group sense) in the low energy regime of interest, where the EFT applies.\\

\color{black}In this broad sense, we only assume the existence of local thermodynamic equilibrium and of "small" (or rather "slow") fluctuations around it, whose dynamics is indeed described by Hydrodynamics. Here, all the questions lie behind the words "small" and "slow" and their precise definition and meaning.\color{black}\\

In its linearised version, hydrodynamics is described by a finite set of \textit{hydrodynamic modes} which can be obtained from the knowledge of the conservation equations and the constitutive relations. These modes display dispersion relations which satisfy the requirement:
\begin{equation}
    \lim_{k \rightarrow 0}\omega_i(k)\,=\,0\,.\label{hi}
\end{equation}
Typical examples are diffusive modes $\omega=-i\,D\,k^2$ and propagating sound modes $\omega=\pm v k-i \Gamma k^2$. More broadly, every hydrodynamic mode obeys a dispersion relation:
\begin{equation}
    \omega^{(i)}(k)\,=\,\sum_{n=1}^{\infty}\,\alpha^{(i)}_n\,k^n\,,\label{series}
\end{equation}
where $\omega\in \mathbb{C}$ and $k \in {\rm I\!R}$. These modes are practically obtained from an eigenvalues equation of the form:
\begin{equation}
    \prod_{j=1}^N\,\left(\omega\,-\,\omega^{(j)}(k)\right)\,\equiv\,F(\omega,k^2)\,=\,0\,,
\end{equation}
which in general not only contains hydrodynamic modes but non-hydrodynamics\footnote{Not satisfying the relation Eq.\eqref{hi}.} ones as well.\\

By staring at Eq.\eqref{series}, one could immediately ask if such a perturbative series is convergent and if yes which is its radius of convergence. This is tantamount to ask what is the radius of convergence of linearised hydrodynamics (in momentum space)\footnote{In this work, we will not consider the question of convergence in real space and for the full non-linear dynamics. For that, see for example \cite{Heller:2020uuy} and specially \cite{Romatschke:2017ejr}.} -- its regime of applicability. Ref.\cite{Withers:2018srf} and later on, in a more complete form, Refs.\cite{PhysRevLett.122.251601,Grozdanov2019} suggested that in order to answer such question, one has to formally extend the function $F(\omega,k^2)$ in complex momentum space and treat it as a complex algebraic curve. In this language, series like Eq.\eqref{series} are known as Puiseux series and their radius of convergence is fundamentally connected to the so-called \textit{critical points} $\{\omega_c,k_c\}$ -- points at which:
\begin{equation}
    F(\omega_c,k_c^2)=0\,,\quad\,\frac{\partial\,F(\omega_c,k_c^2)}{\partial \omega}\,=\,0\,,\label{find}
\end{equation}
for both $\omega_c,k_c^2\,\in\,\mathbb{C}$. These critical points signal ''irregularities'' in the algebraic curve and they are related to         quasinormal modes (QNMs)\footnote{By these we refer simply to all modes with dispersion relation $\omega(k)$ with $k \in {\rm I\!R}$. In other words, QNMs are all the excitations of a specific system \cite{PhysRevD.72.086009}.} crossing. The radius of convergence $\mathcal{R}$ of the hydrodynamic expansion is then set by the distance to the first of these critical points:
\begin{equation}
    \mathcal{R}\,\equiv\,|k_c|\,.
\end{equation}
Moreover, the radius of convergence is intimately connected to the existence of non-hydrodynamic excitations and their interactions with the hydrodynamics modes.\\

This mathematical machinery is extremely elegant but also rather abstract and hard to digest. So far, it has been applied only to few holographic models \cite{Abbasi:2020ykq,Jansen:2020hfd} whose relevance for more realistic situations is at least disputable. In this work, we address the question of the regime of applicability of hydrodynamics in ''real'' liquids and in particularly we ask the following question: given a specific liquid, until which length-scale are we allowed to trust the hydrodynamic approximation? This question can be rephrased as ”which is the physical (and not mathematical) scale setting
the breakdown of the hydrodynamic expansion”. \color{black} A naive answer would be that hydrodynamics is a good description of a liquid as far as the full system can be seen as a continuum and not as a set of molecules/particles interacting with each others. Technically, the failure of hydrodynamics is related to the intrusion of non-hydrodynamic modes which cannot be neglected anymore.\\

Combining the mathematical methods of \cite{PhysRevLett.122.251601,Grozdanov2019} with data from experiments and molecular dynamics (MD) simulations, we will indeed show that the limiting length-scale for the hydrodynamic framework is given by an $\mathcal{O}(1)$ fraction of the inter-molecular distance, given formally by the size of Wigner–Seitz cell $a$. We will also discuss how the regime of applicability depends on the temperature $T$ and on the interactions strength.\begin{figure}[ht]
    \centering
    \includegraphics[width=0.65\linewidth]{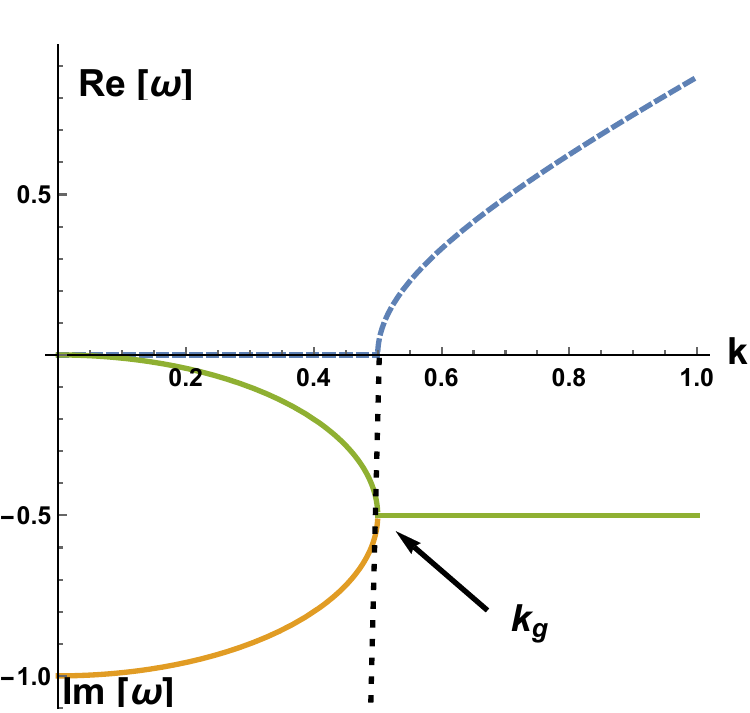}
    \caption{The lowest collective modes in the transverse sector of a liquid, following from the k-gap equation \eqref{kgap}. The real part of the dispersion relation is displayed with dashed line and the imaginary with solid line. We set $v=\tau=1$ for which the cutoff momentum is $k_g=0.5$.}
    \label{fig:0}
\end{figure}\\
As we will discuss in detail, the re-interpretation of the available data will lead us to confirm some intuitive physical arguments but also to new and unexpected findings such as the fact that hydrodynamics \textit{does not} work better \color{black} at strong electromagnetic \color{black} coupling (as always advertised).\color{black}
\subsection*{Convergence of linear hydrodynamics in liquids}
For centuries, the presence of low-frequency propagating shear waves (i.e. transverse phonons) has been considered the fundamental criterion to distinguish solids from liquids. Nevertheless, in the last decade, this definition has been challenged by several experiments observing the presence of shear waves in confined liquids at small frequency \cite{Noirez_2012} and corresponding solid-like elastic effects \cite{Kume2020,kume2020unexpected} (see also \cite{PhysRevLett.62.2616}). This phenomenon indicates that the difference between solids and liquids is only quantitative and it is measured by the relaxation time $\tau$ -- the average time for molecules rearrangement. Within Maxwell theory, such timescale is simply $\tau\equiv \eta/G$ with $\eta$ the shear viscosity and $G$ the shear elastic modulus. It follows that in a solid $\tau=\infty$ -- molecules do not re-arrange but just oscillate around their equilibrium positions.
\begin{figure}[ht]
    \centering
    \includegraphics[width=0.75\linewidth]{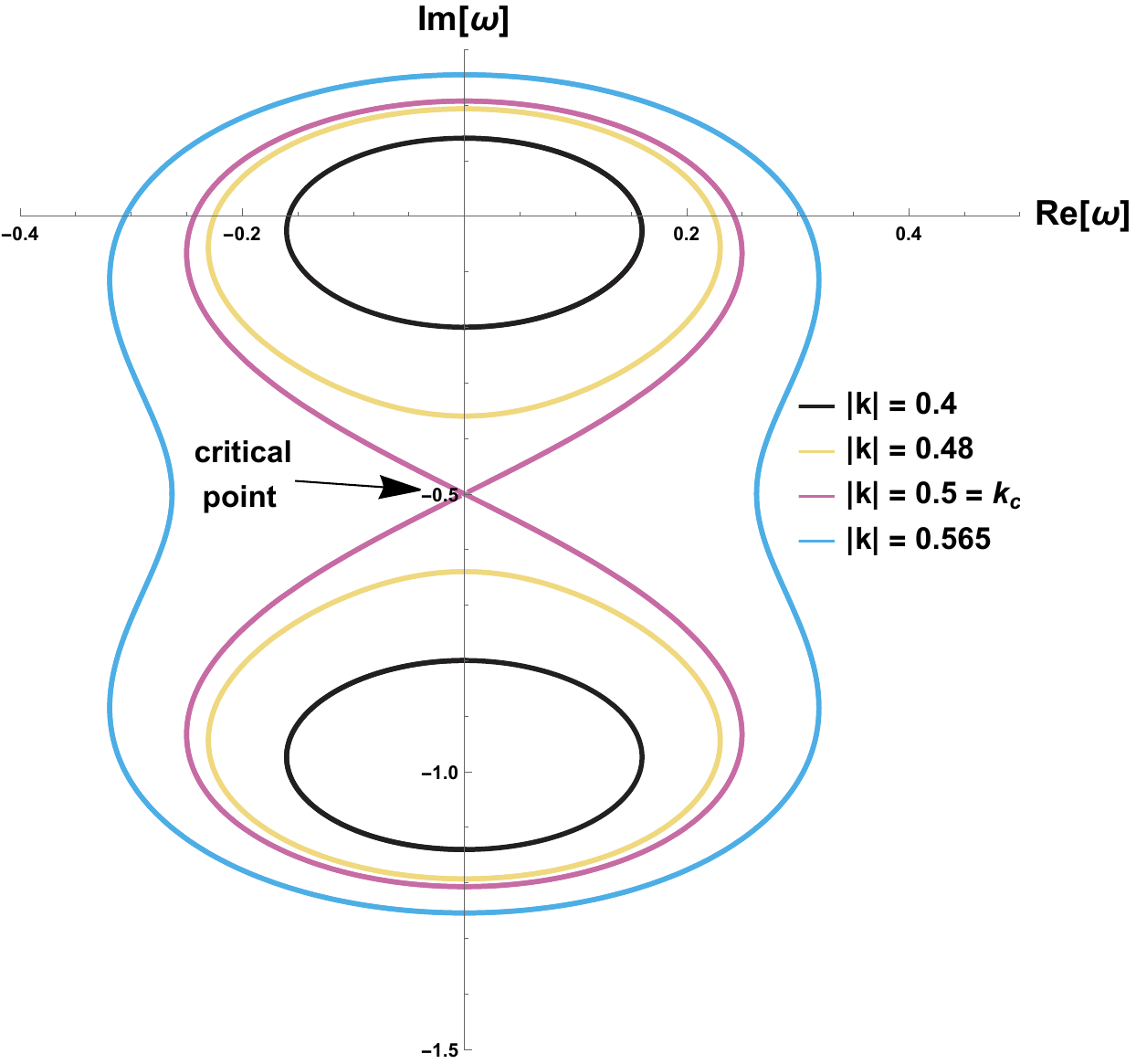}
    \caption{The complex curves corresponding to Eq.\eqref{kgap} for complex momentum $k=|k| e^{i\,\theta}$ with $\theta \in [0,2\pi]$ and varying the modulus $|k|$. The magenta curve corresponds to the critical momentum $|k_c|=0.5$. We set $v=\tau=1$.}
    \label{fig:1}
\end{figure}\\
From a theoretical point of view, this program goes under the name of Maxwell interpolation and its main result is that the dynamics of shear waves in a fluid is governed by the simple equation
\begin{equation}
    \omega^2\,+\,i\,\omega/\tau\,-\,v^2\,k^2\,=\,0\label{kgap}\,,
\end{equation}
which is known for historical reason as the \textit{telegraph equation}. Here, $v$ is the asymptotic speed of the shear waves, dictated by the elastic modulus. Solving Eq.\eqref{kgap}, one finds a couple of modes
\begin{equation}
    \omega\,=\,-\,\frac{i}{2\,\tau}\,\pm\,\sqrt{v^2\,k^2\,-\,\frac{1}{4\,\tau^2}}\,,
\end{equation}
and the existence of a critical momentum $k_g$, labelled as \textit{k-gap}, at which transverse waves start to propagate ($\mathrm{Re}\,\omega \neq 0$). This mechanism appears in several physical contexts; we refer the Reader to \cite{BAGGIOLI20201} for a complete review of them.\\

The appearance of propagating waves for $k>k_g$ is the consequence of the collision between the shear diffusion mode and a non-hydrodynamic excitation, which happens exactly at $k=k_g$ (see Fig.\ref{fig:0}). Given Eq.\eqref{kgap}, it is straightforward to apply the methods of \cite{PhysRevLett.122.251601,Grozdanov2019} and determine the critical point of the corresponding algebraic curve using Eq.\eqref{find}. One immediately obtains that $\omega_c=-i/2 \tau$ and more importantly that
\begin{equation}
       \mathcal{R}\,\equiv\, |k_c|\,=\,\frac{1}{2\,v\,\tau}\,=\,k_g\,.
\end{equation}
This result is confirmed by explicitly computing the set of algebraic curves in the complex plane for complex momentum (see Fig.\ref{fig:1}). Notice that this already implies:
\begin{equation}
   |\omega_c|\,\tau\,=\,\frac{1}{2}\,,
\end{equation}
meaning that linearised hydrodynamics breaks down for timescale shorter than twice the  intrinsic relaxation time.\\

In summary, in realistic liquids, in which the dynamics of shear waves is well described by the k-gap equation \eqref{kgap}, hydrodynamics applies from large distances until a ''microscopic'' length-scale given by the inverse of the k-gap momentum $\mathrm{L}\equiv k_g^{-1}$.\\

Before proceeding, it is important to discuss the assumptions behind the \textit{telegraph equation} \eqref{kgap} and its validity. Clearly, this quadratic equation is a good description of the low energy dynamics only when the system displays a separation of scales between the two lower excitations considered in Eq.\eqref{kgap} and the rest of the modes (i.e. higher order corrections in \eqref{kgap}). This depends a priori on the microscopic details of the system and it can be for instance achieved whenever the relaxation time $\tau$ is parametrically long (e.g. in presence of a weakly broken symmetry) and the quasi-hydrodynamics description of \cite{PhysRevD.99.086012} is valid. Two more comments are in order. (I) In our analysis, we do not rely at any time on the specific relation given by the \textit{telegraph equation} $k_g=1/(2 v \tau)$. Higher momentum corrections obviously modify such expression; nonetheless, the radius of convergence would still be given by the collision point and well approximated by the value at which the real part of the dispersion relation becomes non-zero. (II) The numerical dispersion relations from which the data presented are extracted are very well fitted by the square-root k-gap expression coming from Eq.\eqref{kgap}, at least in the vicinity of the critical momentum of interest. This is an \textit{a posteriori} proof that the potentially dangerous higher order corrections are indeed negligible and that higher modes do not participate in this dynamics. In summary, within the regimes discussed, our analysis is expected to be robust, independent of the validity of the simplex Maxwell model and affected by higher order corrections only in a minor way.\color{black}
\subsection*{Data from experiments and simulations}
\begin{table}
\begin{tabular}{ |c|c|c| }
\hline
 \textbf{Liquid} & $k_c\,a$ \\
\hline
\hline
2D Yukawa (MD) \cite{PhysRevLett.92.065001}&  \, $ 0.25$\,\\
\hline
\hline
dusty plasma (MD) \cite{PhysRevLett.85.2514,PhysRevLett.84.6026} &  \,0.3-1.2\,\\
\hline
\hline
2D Yukawa (EXP) \cite{PhysRevLett.97.115001} & 0.16-0.31 \\
\hline
\hline
Liquid Fe (EXP) \cite{Hosokawa_2015,doi:10.1063/1.5088141}&   0.3\\
\hline
\hline
Liquid Cu (EXP) \cite{Hosokawa_2015,doi:10.1063/1.5088141}&   0.4\\
\hline
\hline
Liquid Zn (EXP) \cite{Hosokawa_2015,doi:10.1063/1.5088141}&   0.3\\
\hline
\hline
3D LJ fluid (MD) \cite{PhysRevLett.125.125501}&   0.2-0.7 \\
\hline
\hline
Liquid Fe (MD) \cite{PhysRevLett.125.125501}&  0.2-0.7 \\
\hline
\hline
IPL8-IPL12 fluid (MD) \cite{PhysRevLett.125.125501}&    0.2-0.7\\
\hline
\hline
Liquid Hg (MD) \cite{PhysRevLett.125.125501}&  0.15-0.55 \\
\hline
\hline
Supercritical Ar (MD) \cite{PhysRevLett.118.215502}&   0.05-0.8\\
\hline
\hline
Subcritical liquid Ar (MD) \cite{PhysRevLett.118.215502}&   0.2-0.7\\
\hline
\hline
Supercritical CO$_2$  \cite{PhysRevLett.118.215502}&   0.1-0.5\\
\hline
\hline
Liquid Ga (EXP, MD) \cite{PhysRevB.101.214312}& 0.25-0.6\\
\hline
\hline
2D Coulomb classical fluids (MD) \cite{doi:10.1063/1.5050708}&  0.3-2 \\
\hline
\hline
Quark Gluon Plasma \cite{Romatschke:2016hle} & 3.3\\
\hline
\end{tabular} 
\caption{A summary of the available data for the momentum cutoff of shear waves $k_c$ in liquids and plasma. EXP stands for experiments while MD for molecular dynamics simulation. The inter-atomic distance is defined as $a$.}
\label{tab1}
\end{table}
\begin{figure}[ht]
    \centering
    \includegraphics[width=1\linewidth]{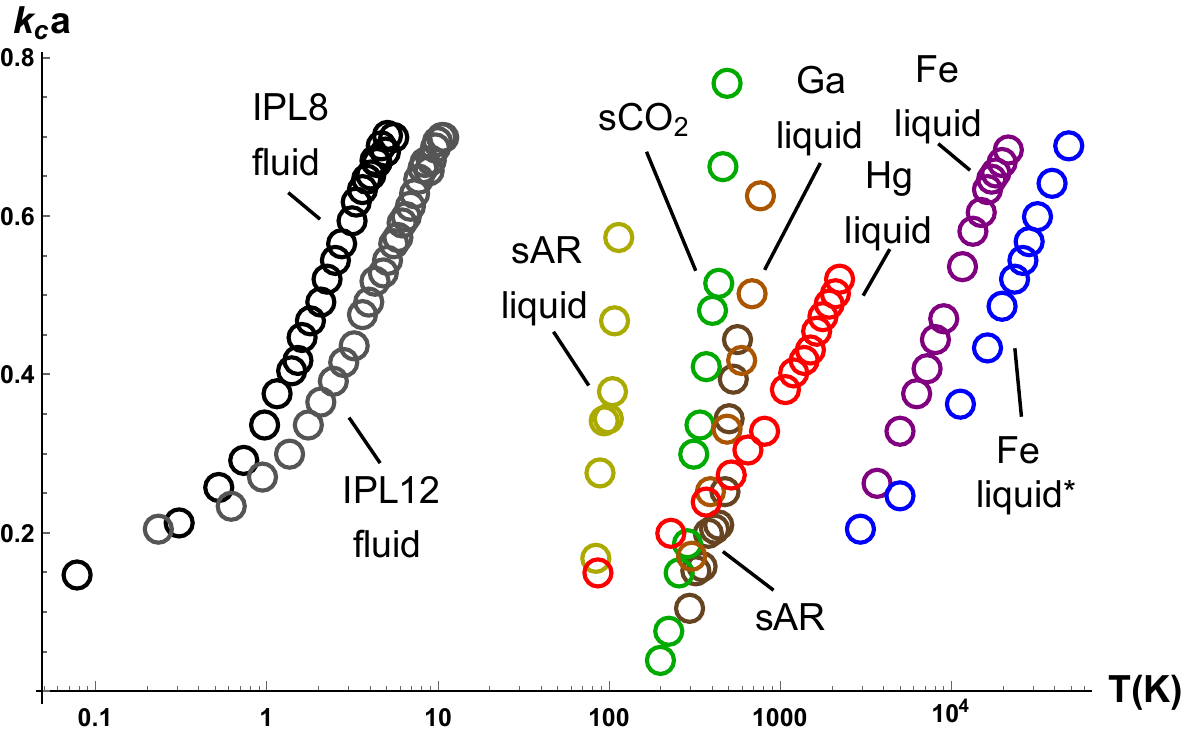}
    \caption{The dimensionless combination $k_c\,a$ with $k_c=k_g$ determining the radius of convergence and $a$ the intermolecular characteristic distance. Temperature dependence for IPL8 fluid (black) \cite{PhysRevLett.125.125501}, IPL12 fluid (gray) \cite{PhysRevLett.125.125501}, liquid supercritical Argon sAr (yellow) \cite{PhysRevLett.118.215502}, supercritical CO$_2$ (green) \cite{PhysRevLett.118.215502}, supercritical argon sAR (brown) \cite{PhysRevLett.118.215502}, liquid Ga (orange) \cite{PhysRevB.101.214312}, liquid Hg (red) \cite{PhysRevLett.125.125501}, liquid Fe $\rho=8$ gm/cm$^3$ (purple) \cite{PhysRevLett.125.125501} and liquid Fe $\rho=10$ g/cm$^3$ (blue) \cite{PhysRevLett.125.125501}.}
    \label{fig:2}
\end{figure}
The validity of the k-gap equation \eqref{kgap} to describe the shear dynamics in liquids has been extensively corroborated by several molecular dynamics simulations (MD) and experiments (EXP) in 2D and 3D liquids and plasmas \cite{PhysRevLett.92.065001,PhysRevLett.85.2514,PhysRevLett.84.6026,PhysRevLett.97.115001,Hosokawa_2015,doi:10.1063/1.5088141,PhysRevLett.125.125501,PhysRevLett.118.215502,PhysRevB.101.214312,doi:10.1063/1.5050708}. The dispersion relation of the transverse waves is usually extracted from the transverse current correlation function $\langle J(-k,t) J(k,0)\rangle$ and within the numerical/experimental errors is well fitted by
\begin{equation}
    \mathrm{Re}\,\omega=\sqrt{v^2\,k^2\,-\,\frac{1}{4\,\tau^2}}\,\quad \text{for}\quad k>k_g .
\end{equation}
It is therefore obvious how to extract the critical momentum from the data.\\

Unfortunately, the critical point determining the radius of convergence in the longitudinal sector is not located at real values of the momentum $k$. For this reason, we are not in the position to complete this analysis including the longitudinal sector. To the best of our knowledge, it is at the moment impossible to obtain data for complex momentum (and complex frequency). Moreover, a complete understanding of the longitudinal sound dynamics in liquids (e.g. the ''positive sound dispersion'' phenomenon \cite{PhysRevA.34.602}), analogous to the k-gap phenomenon for shear waves, is still elusive. A priori, one cannot determine whether the constraint on convergence coming from the longitudinal sector would be more stringent than the one derived in this work. Nevertheless, from a simple physical argument, one would expect both scales to be controlled by the only microscopic length scale in liquids -- the inter-molecular distance.\color{black}\\

A fundamental parameter, both in the simulations and in the experiments, is given by the inter-molecular distance
$a$, which is extracted from the position of the first maximum in the pair distribution function. Technically, this scale corresponds to the Wigner–Seitz radius $a=\left(\frac{2(D-1)}{D}\,n\right)^{-1/D}$, where $n$ is the number density and $D$ the number of spatial dimensions.\\
In Table \ref{tab1}, we list the value of the product $k_c a$ for a large number of 2D and 3D liquids and plasma and, for all the systems considered, we do find that:
\begin{equation}
    k_c\,a\,\approx\,\mathcal{O}\left(1\right)\,.\label{rel}
\end{equation}
This result robustly indicates that the radius of convergence of hydrodynamics is set by the inter-molecular distance\footnote{In principle, the critical scale $k_c$ could be pushed all the way down to zero by achieving a very large relaxation time $\tau$ (i.e. in glasses or solids). Nevertheless, in that situation, the hydrodynamic window $\omega\,\tau\,\ll 1$ would shrink completely and therefore hydrodynamics would not be applicable at all. We thank Kostya Trachenko for pointing this out.}. Moreover, it is in agreement with the naive idea that hydrodynamics applies up to the scale at which you can resolve that the continuum of the liquid is actually formed by a collection of particles. It is important to notice that the relation \eqref{rel} is highly non trivial since the Wigner–Seitz radius can vary from ${\AA}$ ($10^{-10}$ m) for most of the fluids to mm scale in dusty plasmas (see for example \cite{PhysRevLett.97.115001}).\\

Interestingly, the critical momentum for Quark Gluon Plasma was estimated in \cite{Romatschke:2016hle} around $k_c^{-1}\approx 0.15 $ fm. Assuming a inter-parton distance of the order of $a\approx 0.5$ fm \cite{RevModPhys.89.035001}, we again obtain a value $k_c\,a=\mathcal{O}(1)$. This fact is reminiscent of the universality found in the values of the kinematic viscosity in \cite{Baggioli:2020lcf}. \color{black} As a comment, it would be interesting to understand which quantity plays the role of the microscopic inter-molecular scale $a$ in a liquid which appears to be strongly coupled at all energy scales (if it exists).\color{black}\\

Given the abundance of available data, we can do one step further and investigate the temperature dependence of the hydrodynamic convergence radius. In Fig.\ref{fig:2} we have plotted the data for nine different liquids in a large range of temperatures. In all the cases, the regime of applicability increases with temperature and it is consistent with the idea that $k_g \sim 1/v \tau$ and $\tau$ decreases monotonically with temperature (in standard liquids following the well-known Arrhenius law). It also connects with the idea that at $T=0$ (or equivalently $\tau=\infty$) hydrodynamics is not applicable\footnote{This follows simply by the violation of the requirements $\omega/T \ll 1$ or $\omega \tau \ll 1$.}.\\

\begin{figure}[ht]
    \centering
    \includegraphics[width=\linewidth]{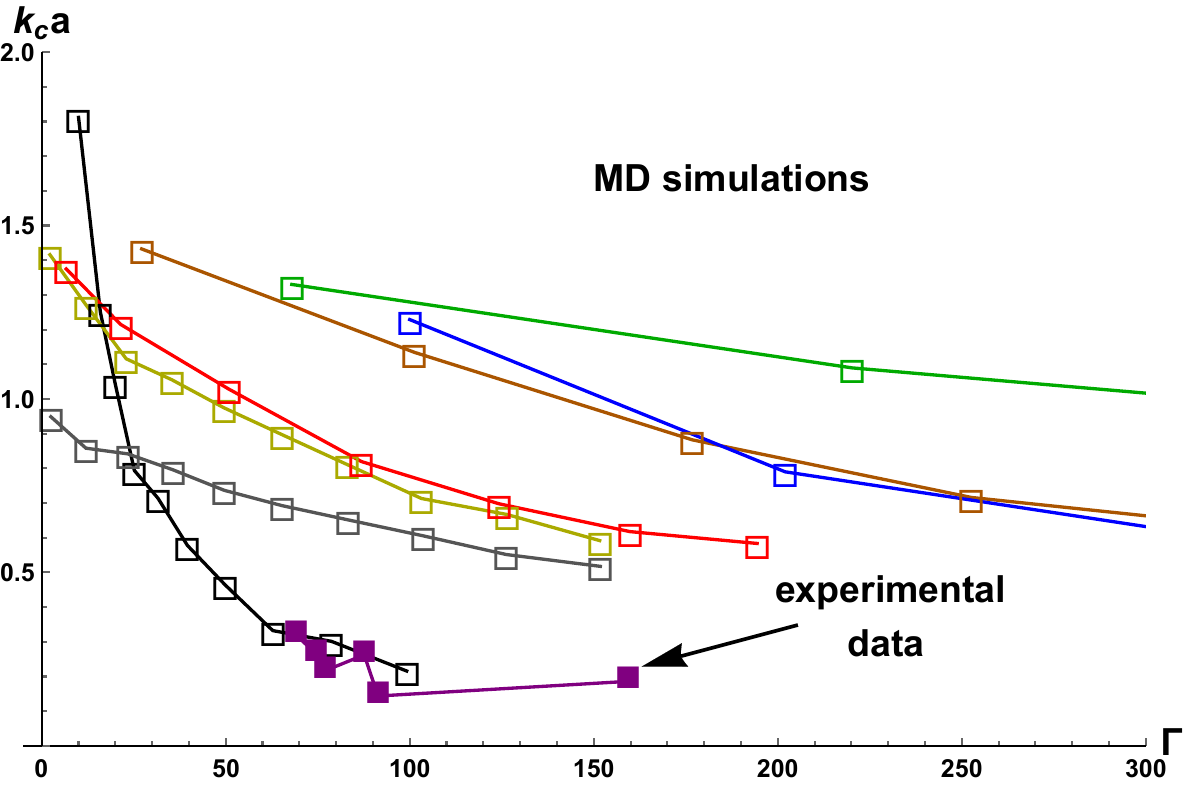}
    \caption{The dimensionless combination $k_c\,a$ with $k_c=k_g$ determining the radius of convergence and $a$ the intermolecular characteristic distance. The empty squares indicate data from MD simulations taken from \cite{doi:10.1063/1.5050708,doi:10.1063/1.5088141,PhysRevLett.85.2514}. The filled squares are experimental data from \cite{PhysRevLett.97.115001}. See also \cite{PhysRevE.85.066401} for a similar collection of data.\color{black}}
    \label{fig:3}
\end{figure}

Finally, using the data for Coulomb fluids and plasmas we can also investigate the radius of convergence in terms of the effective coupling parameter:
\begin{equation}
    \Gamma\,\equiv\,\frac{Q^2}{4\,\pi\,\epsilon_0\,a\,k_B\,T}\,,
\end{equation}
which determines the strength of the Coulomb interactions in the plasma. Here $Q$ is the charge and $  \epsilon_0$ the dielectric constant. The data from simulations universally show (see Fig.\ref{fig:3}) a power-like decrease of the critical momentum in function of $\Gamma$, which is further confirmed by experimental data \cite{PhysRevLett.97.115001}. The dusty plasma considered is a suspension of highly-charged micro-size particles which repel each others via a nearly-Coulomb potential:
\begin{equation}
    U(r)\,=\,\frac{Q}{4\,\pi\,\epsilon_0\,r}\,e^{-r/\lambda_D}\,,
\end{equation}
in which the exponential correction takes into account the screening effects via the screening length $\lambda_D$.
\\
The results therefore demonstrate that stronger Coulomb repulsion decreases the regime of applicability of hydrodynamics. This observation seems to contradict the common wisdom that hydrodynamics works better for strongly coupled systems in which the mean free path becomes smaller (see for example Fig.4 in \cite{Romatschke:2016hle} and the Supplemental material). Nevertheless, it is important to notice that the coupling therein refers to the nuclear interactions -- the strength of the bound states (corresponding to the $SU(N)$ gauge group) -- and not to the inter-molecular Coulomb force\footnote{We thanks Saso Grozdanov for suggesting this point.}. Therefore, no evident discordance is present.
\subsection*{Outlook}
In this work, we have combined the theoretical methods proposed by \cite{PhysRevLett.122.251601,Grozdanov2019} with a large set of data from molecular dynamics simulations and experiments in 2D and 3D liquids and plasmas to determine the regime of applicability of hydrodynamics in realistic systems -- the hydrodynamic convergence radius. Our main result is that the convergence radius is always set in terms of the inter-molecular distance, namely the length-scale at which one can resolve the independent set of particles within the fluid continuum.\\

Moreover, the data indicate that the regime of validity of hydrodynamics increases with the temperature $T$ and decreases with the effective \color{black} electromagnetic \color{black} coupling parameter $\Gamma$. This trend is totally unexpected and against the usual (nevertheless never demonstrated) slogan that ''\textit{hydrodynamics works better at strong coupling}''. This point definitely deserves further investigation.\color{black}\\

Interestingly, the critical momentum for plasmon modes\footnote{Plasmons modes are not QNMs but they are solution of the equation $\epsilon(\omega,k)$ where $\epsilon$ is the dielectric constant. The latter can be taken as our abstract function $F(\omega,k^2)$ in \eqref{find}.} in Dirac fluids has been recently computed in \cite{kiselev2020non} and it reads:
\begin{equation}
    k_c\,=\,\frac{\tau_V}{4\,v\,\tau_{c,1}^2}\,,\label{cc}
\end{equation}
where $\tau_{c,1}$ is the relaxation time, $v$ the electrons group velocity and $\tau_V$ is the time scale characterizing electrostatic interaction. In the near future, comparing better \eqref{cc} with our results would be desirable.\\

Taking into account the fundamental role of non-hydrodynamic modes, it would be extremely helpful to build experimental setups able to pinpoint them on the lines of \cite{PhysRevLett.115.190404}. Moreover, it would be interesting to understand which are the experimental and physical consequences of the non-convergence of the hydrodynamic series in realistic systems\footnote{We thank Egor Kiselev for suggesting this point.}.\\

Finally, the k-gap dynamics \eqref{kgap} universally appears in the context of diffusive Goldstone bosons \cite{Minami:2018oxl,Hidaka:2019irz} in dissipative systems (e.g. in quasicrystals \cite{Currat2002,Baggioli:2020haa}). It would be interesting to connect such mechanism with the results of this manuscript.\\

In conclusion, given the extremely wide usage of hydrodynamic methods, it is mandatory to understand until which length-scale those can be trusted. In this work, we provided a direct and pragmatic answer to this question in a large set of realistic liquids and plasmas. We hope that our results will boost the efforts to deeply understand hydrodynamics and its regime of applicability and to connect more closely theory with experiments.
\subsection*{Acknowledgments} 
The author would like to thank A. Zaccone, S.Yurchenko, L.Noirez, N.Poovuttikul, N.Abbasi, S.Grieninger, M.Heller, J.Schmalian, E.Kiselev, K.Landsteiner and specially S.Grozdanov, P. Romantschke and K.Trachenko for fruitful discussions and insightful comments on the topics covered in this manuscript.
M.B. acknowledges the support of the  Shanghai Municipal Science and Technology Major Project (Grant No.2019SHZDZX01) and
of the Spanish MINECO “Centro de Excelencia Severo Ochoa” Programme under grant
SEV-2012-0249.
\bibliographystyle{apsrev4-1}

\bibliography{real}
\appendix
\subsection*{Holographic examples}
A common playground where the k-gap equation (6) appears is in the context of holographic models with higher-from global symmetries \cite{Grozdanov:2016tdf,SciPostPhys.4.1.005,PhysRevD.99.086012,PhysRevD.97.106005}. These models are used to described both magneto-hydrodynamics in charged plasmas and the elasto-dynamics in viscoelastic media. In the first scenario, the emergent propagating degree of freedom at $k=k_g$ is the photon. In the neutral plasma approximation, the corresponding critical momentum is given in terms of \cite{Grozdanov:2017kyl}
\begin{myequation}
    k_c\,\approx\,\frac{\pi\,T}{2\,\log(M/\pi T)}\,,
\end{myequation}
where $T$ is the temperature and $M$ is an UV cutoff of the theory.\\

In the second scenario, the emergent mode is a propagating phonon and the critical momentum reads \cite{Grozdanov:2016vgg}
\begin{myequation}
\resizebox{.47 \textwidth}{!} {$ k_c\,=\,8 \pi ^2 \sqrt{\frac{T^4}{(3 M-4 \pi  T) \left(27 M-2 \pi  T \left(18+\sqrt{3} \pi -9 \log
   (3)\right)\right)}} \,$,}
\end{myequation}
where the same notations are used.
\begin{figure}[hb]
    \centering
    \includegraphics[width=0.7\linewidth]{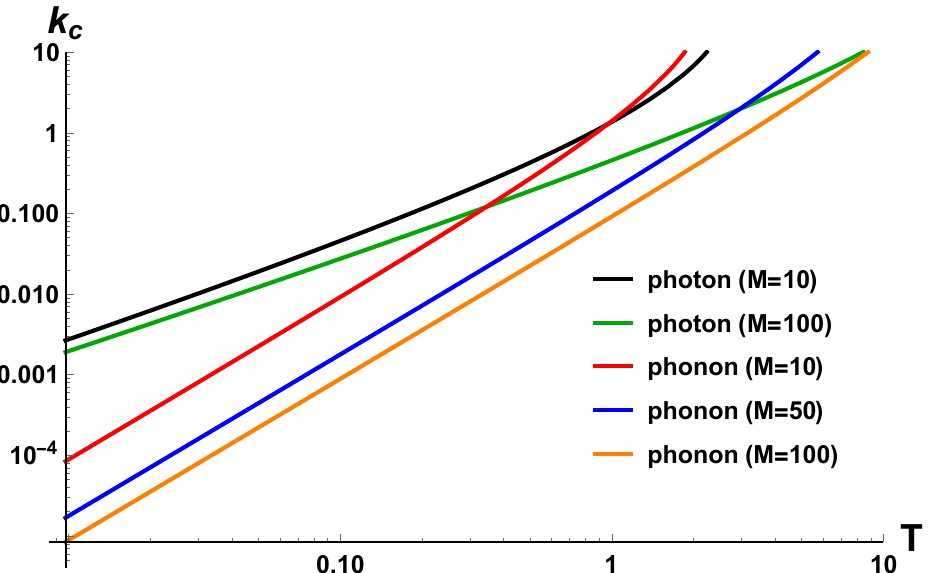}
    \caption{The critical momentum in the higher-form global symmetries holographic models. The formulas are taken from \cite{SciPostPhys.4.1.005,PhysRevD.99.086012} for the photon propagator and from \cite{PhysRevD.97.106005} for the viscoelastic model.}
    \label{fig:app1}
\end{figure}\\
It is simple to see that in both cases the critical momentum, determining the radius of convergence of hydrodynamics, grows monotonically with temperature independently of the value of the UV cutoff $M$.\\

Another situation where the k-gap equation (6) is at work is in models with dynamical Coulomb interactions and emergent plasmonic modes \cite{Gran2018,Baggioli2019,Baggioli2020}. In those models, it is possible to obtain numerically the dispersion relations in function of the electromagnetic coupling $\lambda$. 
\begin{figure}
    \centering
    \includegraphics[width=0.7\linewidth]{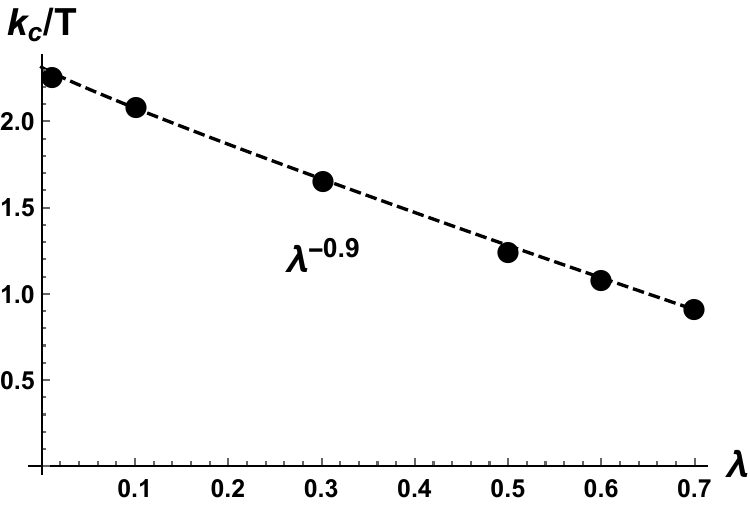}
    \caption{The critical momentum in function of the electromagnetic coupling $\lambda$ in the holographic model of \cite{Baggioli2019}.}
    \label{fig:app2}
\end{figure}
The numerical results are shown in Fig.\ref{fig:app2} and they clearly indicate a power-law fall-off of the critical momentum in function of $\lambda$. This is in agreement with the data obtained from 2D plasmas and presented in the main text (see Fig.4) under identifying $\lambda$ with $\Gamma$.\\

Finally, a well-known place where the k-gap appears is in the Israel-Stewart theory for linearised relativistic hydrodynamics \cite{ISRAEL1979341}. In this model, the critical momentum is given by
\begin{myequation}
    k_c\,=\,\frac{1}{\sqrt{4\,D\,\tau}}\,=\sqrt{\frac{\epsilon+p}{4\,\eta\,\tau_\pi}}\,.
\end{myequation}
\begin{figure}[hb]
    \centering
    \includegraphics[width=0.7\linewidth]{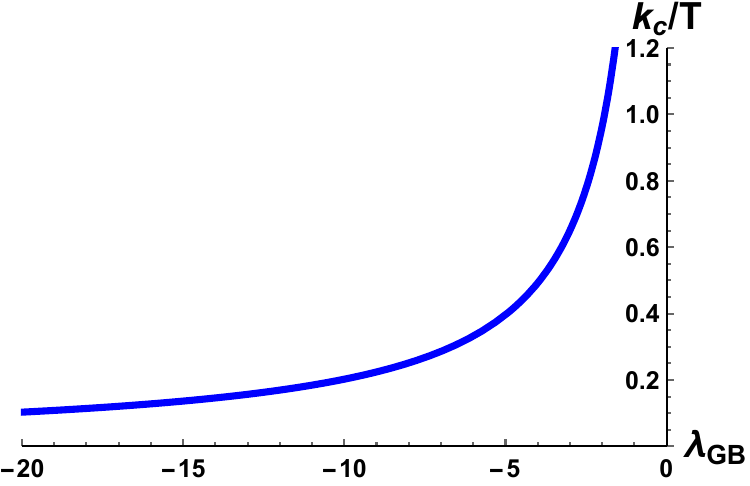}
    \caption{The critical momentum $k_c=1/2 v \tau$ in function of the Gauss Bonnet coupling. Formulas taken from \cite{PhysRevD.99.086012}.}
    \label{fig:app3}
\end{figure}\\
The gravity dual of the Israel-Stewart theory is provided by a Gauss-Bonnet model \cite{PhysRevD.99.086012} ,where the k-gap momentum reads:
\begin{myequation}
  k_c\,=\,  -\frac{\sqrt{2} \pi  T \sqrt{-\frac{\lambda _{\text{GB}}}{\gamma  (\gamma +2)+2 \log \left(\frac{2}{\gamma +1}\right)-3}}}{\lambda _{\text{GB}}}\,,\label{ll}
\end{myequation}
with $\gamma\equiv \sqrt{1-4\,\lambda _{\text{GB}}}$ and $\lambda_{GB}\in [-\infty,1/4]$.\\
From Eq.\eqref{ll} two facts are evident. (I) The critical momentum grows linearly with temperature in agreement with the data shown in Fig.3. (II) The critical momentum decreases by increasing the value of the Gauss-Bonnet coupling on the negative axes. Given that such direction corresponds to decrease the coupling of the dual field theory from infinity, the results shown in Fig.\ref{fig:app3} suggest that hydrodynamics works better at strong coupling. This outcome is in agreement with \cite{Romatschke:2016hle}. Notice that the disagreement with the behaviour with respect to $\Gamma$ is certainly puzzling but it does not constitute a discordance since $\lambda$ refers to the $SU(N)$ coupling while $\Gamma$ to the inter-molecular Coulomb potential.
\subsection*{The telegraph}
\begin{figure}[hb]
    \centering
    \includegraphics[width=0.7\linewidth]{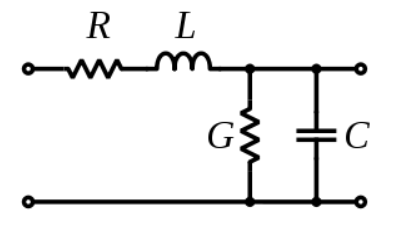}
    \caption{Schematic of the electric circuit giving rise to the telegraph equation Eq.\eqref{teleq}.}
    \label{fig:tele}
\end{figure}

As explained in the main text, Eq.(6) is also known as the \textit{telegrapher equation} and it was first written by Oliver Heaviside in the attempt of studying the dynamics of transmission lines. In particular, such equation comes from the analysis of a specific circuit displayed in Fig.\ref{fig:tele}, where $R,L,G,C$ are respectively a resistance, an inductance, a conductance and a capacitance. In such a configuration, the equation for the voltage $v$ across the piece of wire at position $x$ at time $t$ is given by:
\begin{myequation}
    L\,C\,\frac{\partial^2 v}{\partial t^2}\,+\,\left(L\,C\,+\,R\,G\right)\,\frac{\partial v}{\partial t}\,+\,R\,G\,v\,=\,\frac{\partial^2 v}{\partial x^2}\label{teleq}\,,
\end{myequation}
which is a generalized form of Eq.(6) with the presence of an extra mass term (vanishing for either $R=0$ or $G=0$). In any case, the solution enjoys very similar properties and in particular the presence of a critical momentum defined as
\begin{equation}
    k_c\,=\,\frac{|G\,L\,-\,C\,R|}{2\,\sqrt{C\,L}}\,.
\end{equation}
Interestingly, using some data of telegraph wires for transmissions at low frequency reported in \cite{chen2004home}, one obtains:
\begin{myequation}
    k_c\,\approx\,0.79 \,\text{Km}^{-1}\,,
\end{myequation}
which means that the radius of convergence of hydrodynamics in a telegraph wire is about $L\approx 1.26$ Km! Moreover, one could easily extract the relaxation time:
\begin{myequation}
    \tau\,=\,\frac{C\,L}{G\,L\,+\,C\,R}\,\approx\,10^{-6}\,s\,,
\end{myequation}
which determines the timescale at which hydro fails. Notice that obviously:
\begin{myequation}
    \frac{1}{k_c\,\tau}\,\approx\,3\,\times\,10^{8}\,\text{m/s}\,\equiv\,c\,,
\end{myequation}
i.e. the speed of light -- the speed of propagation of electromagnetic waves inside the transmission wire.
\end{document}